\documentclass[twocolumn,english,aps,prl,showpacs,superscriptaddress,preprintnumbers]{revtex4}
\usepackage[T1]{fontenc}
\usepackage[latin9]{inputenc}
\usepackage{float}
\usepackage{amsmath}
\usepackage{amssymb}
\usepackage{graphicx}
\usepackage{esint}
\usepackage{units}

\makeatletter

\usepackage{amstext}

\usepackage{babel}

\addto\captionsenglish{}
\makeatother

\usepackage{babel}
\begin{document}

\title{Manipulating the Interaction between Localized and Delocalized Surface Plasmon Polaritons in Graphene}

\author{Renwen Yu}

\affiliation{Institute of Condensed Matter Theory and Solid State Optics, Abbe
Center of Photonics, Friedrich-Schiller-Universit\"{a}t Jena, D-07743
Jena, Germany}

\author{Rasoul Alaee}
\affiliation{Institute of Theoretical Solid State Physics, Karlsruhe Institute of Technology, Wolfgang-Gaede-Strasse 1, 76131
Karlsruhe, Germany}

\author{Falk Lederer}

\affiliation{Institute of Condensed Matter Theory and Solid State Optics, Abbe
Center of Photonics, Friedrich-Schiller-Universit\"{a}t Jena, D-07743
Jena, Germany}

\author{Carsten Rockstuhl}
\affiliation{Institute of Theoretical Solid State Physics, Karlsruhe Institute of Technology, Wolfgang-Gaede-Strasse 1, 76131 Karlsruhe, Germany}
\affiliation{Institute of Nanotechnology, Karlsruhe Institute of Technology, P.O. Box 3640, 76021 Karlsruhe, Germany}

\begin{abstract}

The excitation of localized or delocalized surface plasmon polaritons in nanostructured or extended graphene has attracted a steadily increasing attention due to their promising applications in sensors, switches, and filters. These single resonances may couple and intriguing spectral signatures can be achieved by exploiting the entailing hybridization. Whereas thus far only the coupling between localized {\it or} delocalized surface plasmon polaritons has been studied in graphene nanostructures,  we consider here the interaction between a localized {\it and} a delocalized surface plasmon polariton. This interaction can be achieved by two different schemes that reside on either evanescent near- field coupling or far-field interference. All observable phenomena are corroborated by analytical considerations, providing insight into the physics and paving the way for compact and tunable optical components at infrared and terahertz frequencies.
\end{abstract}

\pacs{78.67.Wj,
78.20.-e,
73.21.-b,
71.70.Gm
}

\maketitle

\section{Introduction}

Graphene, a two-dimensional (2D) arrangement of carbon atoms, enables a multitude of exciting applications due to its extraordinary electrical, mechanical, but especially, optical properties \cite{novoselov2004electric,geim2007rise,bonaccorso2010graphene,grigorenko2012graphene,tassin2013graphene,iorsh2013hyperbolic}.
The most prominent optical peculiarity is likely its non-dispersive (in a broad spectral domain) absorption of about $2.3\%$ of an incident electromagnetic wave \cite{nair2008fine}. This might sound marginal but is actually exceptionally large, considering that a monolayer of a material with negligible thickness is responsible for such observation.

But this absorption can be even considerably enlarged in a narrow frequency range.  This is possible by exploiting either delocalized (DSPPs) or localized
surface plasmon polaritons (LSPPs). There, the electromagnetic field is resonantly coupled to oscillations of the surface charge density leading to a strong enhancement of light-matter-interaction at the nanoscale \cite{koppens2011graphene}.
The excitation of either graphene DSPPs or LSPPs at IR and THz frequencies has been already demonstrated while relying on various graphene micro- or nanostructures \cite{jablan2009plasmonics,vakil2011transformation,christensen2011graphene,gao2012excitation,nikitin2012surface,farhat2013exciting,buslaev2013plasmons}.
They have been employed to achieve novel exciting
functionalities, such as perfect absorbers \cite{alaee2012perfect,pirruccio2013coherent,thongrattanasiri2012complete},
broadband polarizers \cite{bao2011broadband,cheng2013dynamicallyPolarizor},
THz reflectarrays \cite{carrasco2013tunable}, tunable THz cloaking
\cite{chen2011atomically,chen2013nanostructured}, or tunable graphene
antennas \cite{filter2013tunable}.

However, not just individual resonances sustained by micro- or nanostructured graphene have been studied, but also the excitation of multiples thereof. This can be either done with suitably designed individual elements that exhibit multiple resonances or while coupling multiple elements  with single resonances. Effects like plasmon hybridization \cite{prodan2003hybridization} or spectral interference give rise to  involved but intriguing spectral signatures. They are extremely appealing for a larger variety of applications such as plasmon-induced transparency (PIT) \cite{zhang2008plasmon,liu2009plasmonic}, tunable Fano resonance sensing \cite{hao2008symmetry}, and strong mode confinement \cite{oulton2008hybrid,fedotov2007sharp}.

Recently, PIT has also been demonstrated in
periodically patterned graphene nanostrips. This was possible by exploiting the spectral interference
between a bright and a dark LSPP mode sustained by periodically patterned graphene nanostrips
\cite{cheng2013dynamicallyEIT}. Alternatively, a double graphene layer structure
acting as a plasmonic waveguide switch has been proposed that exploits the culling between two DSPP modes supported by those two graphene layers
\cite{iizuka2013deep}.

However, thus far only the coupling
between graphene DSPPs or LSPPs has been considered, but never that between a DSPP and a LSPP. With the purpose to understand the observable phenomena in such regime, we study here comprehensively this important phenomenon. Emphasis is put on the exploration of two different schemes  where interaction relies either on near-field coupling or far-field interference.

The first scheme exploits a graphene-only device where a LSPP, supported
by a periodic array of graphene ribbons (PAGR), and a DSPP, supported
by an extended graphene layer, are coupled. This coupling is evoked by the near-field overlap of both SPPs if the PAGR and the graphene layer are very close to each other. The strength of the coupling can get very large by carefully tuning the geometrical
parameters of the system. Consequently, the formation of hybrid DSPP-LSPP
polaritons can be witnessed.

The second scheme allows to study the interference between LSPPs and
DSPPs in a far-field interaction regime, facilitated by a dielectric
diffractive grating below the bottom graphene layer. In this regime, a pronounced
Fabry-Perot (FP) effect needs to be taken into account. A simple picture
considering the entire structure as an effective FP cavity is proposed to provide an intuitive understanding
of the far-field interference phenomenon. The numerical results for both interaction schemes are corroborated by analytical models. They yield clear insights into the intriguing optical phenomena observed.

\section{Near-field coupling \textbf{scheme} (NFCS)}

\begin{figure}
\begin{centering}
\includegraphics[width=64mm]{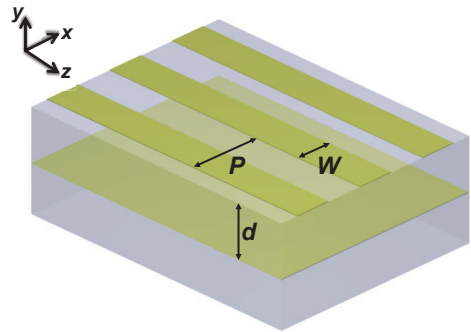}
\par\end{centering}

\caption{Schematic view of the near-field coupling scheme. The PAGR (array
period $P$ and ribbon width $W$) and the bottom graphene layer are
separated by a dielectric spacer of thickness $d$.\label{fig:Schematic_Near}}
\end{figure}

A schematic view of the proposed near-field coupling scheme (NFCS) is displayed
in Fig.\,\ref{fig:Schematic_Near}. The entire structure is
assumed to be imbedded in an ambient material with a relative permittivity of $\varepsilon=2.25$. This, however, is by no means a limitation and any other dielectric material environment could have been considered as well. The
PAGR on top (array period $P$, ribbon width $W$) and the bottom
graphene layer are separated by a dielectric spacer of thickness $d$.
Throughout this work, we assume that the polaritons are excited by a normally
incident TM-polarized wave propagating in negative
$y$ direction with magnetic field polarized along $z$ direction.
Then, the optical response of the PAGR is characterized
by the LSPP resonance \cite{nikitin2012surface,nikitin2011edge}.
Moreover, simultaneously, its periodicity leads to the excitation of
the DSPP modes in the nearby graphene layer due to Bragg diffraction
\cite{gao2012excitation,christ2006controlling}.  Due to the evanescent coupling the excitation efficiency of the DSPP modes in the bottom graphene layer depends strongly
on the separation distance $d$ to the PAGR \cite{gao2012excitation}, as we will show in the
following. The surface conductivity of graphene is modelled within
the local random phase approximation including the finite temperature correction
as \cite{koppens2011graphene}

\begin{equation}
\sigma\left(\omega\right)=\frac{e^{2}E_{\mathrm{F}}}{\pi\hbar^{2}}\frac{i}{\omega+i\tau^{-1}}.\label{eq:simple_con}
\end{equation}

Here, $E_{\mathrm{F}}$ is the Fermi energy, $\tau=\mu E_{\mathrm{F}}/(ev_{\mathrm{F}}^{2})$
is the relaxation time, $\hbar$ is the reduced Planck constant, $e$
is the electron charge, $v_{\mathrm{F}}\approx10^{6}\,\mathrm{m/s}$
is the Fermi velocity, and $\mu\approx10000\,\mathrm{cm^2/V\cdot s}$
is the impurity-limited DC mobility \cite{novoselov2004electric}. All numerical simulations were performed by using a finite
element method (FEM) based solver for Maxwell's equations (\textit{COMSOL Multiphysics}).

\begin{figure}
\begin{centering}
\includegraphics[width=84mm]{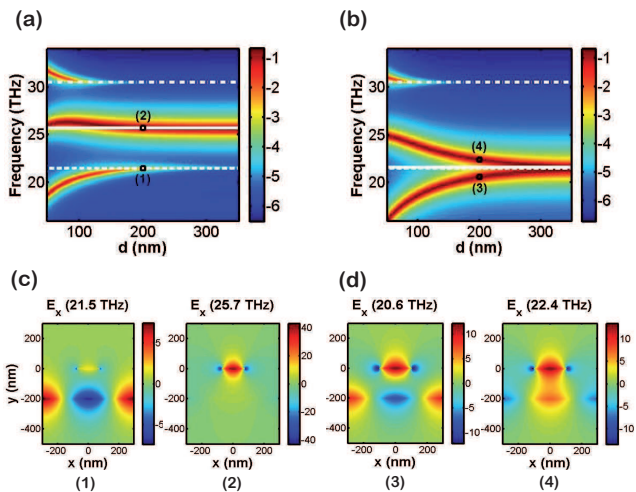}
\par\end{centering}

\caption{(a) and (b): Simulated absorption spectra for two different ribbon
widths: (a) $W=150\,\mathrm{nm}$; (b) $W=210\,\mathrm{nm}$. The
spectral positions of the bare modes are indicated by white horizontal lines
(solid line: LSPP mode, dashed line: first order DSPP mode, dashed-dotted
line: second order DSPP mode). The spectra are plotted in
a logarithmic scale. (c) and (d): Electric field distribution $E_{x}$
of the absorption peaks corresponding to the positions indicated by
(1), (2), (3) and (4) in panels (a) and (b), respectively. Note that
the upper PAGR and the graphene layer are located at $y=0$ and
$y=-200\,\mathrm{nm}$, respectively. All local electric field distributions
are normalized to the incident field.\label{fig:W_Near}}
\end{figure}

To start the analysis a grating period $P=600\,\mathrm{nm}$ has been chosen.
The Fermi energy level $E_{\mathrm{F}}=600\,\mathrm{meV}$ is
the same in the upper PAGR and the bottom graphene layer.  The simulated
absorption spectra are shown in Figs.\,\ref{fig:W_Near}(a) and (b) for two different ribbon widths $W$ as a function of the spacer thickness in the interval from $d=50\,\mathrm{nm}$ to $d=350\,\mathrm{nm}$. The white dashed and dash-doted lines indicate the spectral positions of the DSPPs of the isolated graphene layer excited by the first and second Bragg diffraction order, respectively. They are calculated from the following equation \cite{gao2012excitation,farhat2013exciting}

\begin{equation}
\omega_{\mathrm{D}}=\sqrt{\frac{2Ne^{2}E_{\mathrm{F}}}{\hbar^{2}\varepsilon_{0}\left(\varepsilon_{1}+\varepsilon_{2}\right)P}},\label{eq:f_DSP}
\end{equation}
where $\varepsilon_{0}$ is the vacuum permittivity, $\varepsilon_{1}$
and $\varepsilon_{2}$ are the permittivities of the materials above
and below the graphene layer, and $N$ is an integer denoting
the Bragg diffraction order.
The spectral position of the bare LSPP mode, indicated by the white solid line, is numerically calculated from one isolated PAGR with identical Fermi energy and geometrical parameters.

In Fig.\,\ref{fig:W_Near}(a), for $W=150\,\mathrm{nm}$ and for a large
spacer thickness of $d=350\,\mathrm{nm}$, only a single but
strong absorption peak emerges at around $25.6\,\mathrm{THz}$. This absorption peak
is related to the LSPP resonance, indicated by the white solid line. The mode shape is depicted in Fig.\,\ref{fig:W_Near}(c) and changes only slightly when $d$ decreases down to $100\,\mathrm{nm}$.

When the separation distance decreases to about $d=220\,\mathrm{nm}$,
another smaller absorption peak emerges at around $21.5\,\mathrm{THz}$. It
is associated with the first Bragg diffraction order ($N=1$) and constitutes a hybrid DSPP-LSPP resonance emerging from the bare DSPP mode supported
by the graphene layer. The redshift of this hybrid mode
and the blueshift of the LSPP resonance with decreasing $d$ arises
from the enhanced mutual near-field interaction and is a signature of the small but notable hybridization.

When the separation
distance is further decreased to about $d=120\,\mathrm{nm}$, a
third absorption peak appears at higher frequency (around $30.5\,\mathrm{THz}$).
It is related to another hybrid DSSP-LSSP mode branching off the second order ($N=2$) DSPP resonance. The gradual appearance
of the hybrid modes essentially supported by the bottom graphene layer is due to
the increasing excitation efficiency through the upper PAGR when
the spacer thickness $d$ decreases because of the larger modal overlap between the LSPP and the DSPP. The higher the diffraction order from the PAGR that is used to excite the DSPP, the shorter is the decay length of the associated evanescent wave. Therefore a notably amplitude necessary for the excitation only exists below a specific spacer thickness, and this distance gets smaller the larger the diffraction order is. At the smallest considered spacer thickness of $d=50\,\mathrm{nm}$, a large enhancement of the resonance strength of the DSPP-originated hybrid mode resonances at $18.7\,\mathrm{THz}$
and $31.7\,\mathrm{THz}$ can be recognized. This indicates a strong near field coupling
and energy exchange between the LSPP and DSPP modes in graphene. Moreover, after interacting with the second order DSPP mode, the LSPP mode
slightly redshifts, accompanied by a blueshift of the second
order DSPP mode.

More dramatic variations of the optical response
can be recognized in Fig.\,\ref{fig:W_Near}(b). There,
the ribbon width was slightly increased to $W=210\,\mathrm{nm}$ to achieve a spectral coincidence
of the bare LSPP mode with the bare first order DSPP mode at about $21.5\,\mathrm{THz}$.
As a result a bonding [Fig.\,\ref{fig:W_Near}(d4)] and an anti-bonding [Fig.\,\ref{fig:W_Near}(d3)] hybrid DSPP-LSPP mode occur at smaller
spacer thicknesses $d$. When increasing $d$, those two polariton
modes degenerate.

In order to show that the coupling strength gets enhanced when the spectral
positions of the LSPP mode ($\omega_{\mathrm{L}}$) and the first
order DSPP mode ($\omega_{\text{\ensuremath{\mathrm{D}}}}$) coincide,
the electric field distributions at a fixed separation distance
$d=200\,\mathrm{nm}$ are depicted in Figs.\,\ref{fig:W_Near}(c)
and (d). It can be seen that when $\omega_{\mathrm{L}}\neq\omega_{\mathrm{D}}$,
the electric field is mainly either concentrated in the bottom graphene
layer or in the upper PAGR, as shown in Fig.\,\ref{fig:W_Near}(c). The field distributions
of these weakly hybrid modes resemble either the bare LSPP
mode or the bare first order DSPP mode \cite{gao2012excitation}.
However, when $\omega_{\mathrm{L}}\approx\omega_{\mathrm{D}}$, the
electric field is concentrated in both upper PAGR and bottom graphene
layer as shown in Fig.\,\ref{fig:W_Near}(d). This indicates an enhanced
mode mixing and a larger coupling strength, and also demonstrates
the formation of genuine bonding and anti-bonding LSPP-DSPP polaritons. In addition to varying the ribbon width $W$ as we have shown in Fig.\,\ref{fig:W_Near}, the coupling between the LSPP mode and the second order DSPP mode can also be controlled by properly adjusting the array period $P$.

\begin{figure}
\begin{centering}
\includegraphics[width=84mm]{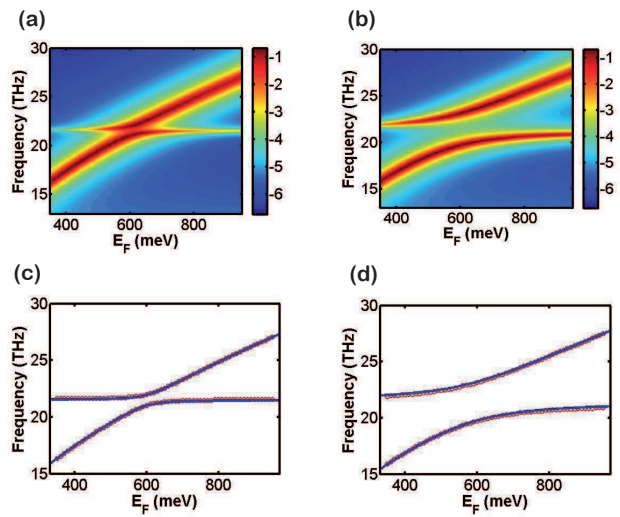}
\par\end{centering}

\caption{(a) and (b): Simulated absorption spectra as function of the Fermi energy at two different separation
distances: (a) $d=260\,\mathrm{nm}$; (b) $d=140\,\mathrm{nm}$. The spectra are plotted in a logarithmic scale. (c) and (d): Positions
of the absorption peaks (red circles) extracted from panels (a) and
(b) as a function of the Fermi energy $E_{\mathrm{F}}$. The blue
solid curves represent the analytical results from Eq.\,\ref{eq:polariton_branch}.\label{fig:Fermi_Near}}
\end{figure}

The Fermi energy
of graphene can be dynamically tuned by electrical gating or chemical doping \cite{novoselov2004electric}. Thus, a similar
tunability can also be expected due to the dynamical manipulation
of the coupling between the LSPP and DSPP modes. Here, we assume
that the geometric parameters of the upper PAGR are fixed as $W=210\,\mathrm{nm}$
and $P=600\,\mathrm{nm}$. The Fermi energy $E_{\mathrm{F}}=600\,\mathrm{meV}$
is constant in the bottom graphene layer, but is subject to changes in the upper PAGR to investigate the tunability inside our NFCS.

As already shown in Fig.\,\ref{fig:W_Near} the coupling between
the DSPP and the LSPP modes can strongly enhance the DSPP resonance.
Thus, the energy exchange between the LSPP and DSPP modes is enhanced
as well, given that the DSPP mode and the LSPP mode coincide spectrally. This leads to a splitting of the absorption peak at the
spectral position of the DSPP mode, which is fixed in this case. Moreover,
we observe that the coupling strength can be modified by tuning
the Fermi energy $E_{\mathrm{F}}$ of the upper PAGR. The variation of
the induced polariton modes with different Fermi energy is displayed
in Fig.\,\ref{fig:Fermi_Near}. Simulated absorption spectra
with two different separation distances $d$ are displayed in Figs.\,\ref{fig:Fermi_Near}(a)
and (b). It  can be see that as $E_{\mathrm{F}}$ is varied, an obvious
anticrossing feature is observed in both Figs.\,\ref{fig:Fermi_Near}(a)
and (b). At smaller $E_{\mathrm{F}}$, the lower polariton behaves LSPP-like
with a broad spectral width. When $E_{\mathrm{F}}$ is tuned
towards the DSPP mode, this polariton mode changes its nature and
gradually DSPP-like features emerge such as with narrow spectral width. The upper
polariton mode follows the opposite route.

In Figs.\,\ref{fig:Fermi_Near}(c)
and (d), the spectral positions of the absorption maxima extracted
from Figs.\,\ref{fig:Fermi_Near}(a) and (b) are depicted as red
circles, which indicate the induced upper and lower LSPP-DSPP polariton
modes ($\omega_{\mathrm{u}}$ and $\omega_{\mathrm{l}}$). At a Fermi
energy $E_{\mathrm{F}}=600\,\mathrm{meV}$, the two polariton modes
anticross. The Rabi splitting $2\hbar\omega_{\delta}$ is $3.9\,\mathrm{meV}$
and $14.3\,\mathrm{meV}$ for the cases of the two separation distance
$d=260\,\mathrm{nm}$ and $d=140\,\mathrm{nm}$, respectively. This
indicates the coupling strength of the LSPP-DSPP interaction in a
sense that the Rabi splitting becomes larger with decreasing the separation
distance $d$ since the interaction between the LSPP and DSPP modes
is stronger. The blue solid curves are the results of solving the
eigenvalue problem

\begin{equation}
\hat{H}\Psi=\hbar\omega_{\mathrm{u,l}}\Psi,\label{eq:eigenvalue_problem}
\end{equation}
with eigenvector $\Psi$ and interaction Hamiltonian $\hat{H}=\hbar\left(\begin{array}{cc}
\omega_{\mathrm{D}} & \omega_{\delta}\\
\omega_{\delta} & \omega_{\mathrm{L}}\left(E_{\mathrm{F}}\right)
\end{array}\right)$. Here, the eigenenergy of the DSPP mode is $\hbar\omega_{\mathrm{D}}=88.9\,\mathrm{meV}$
and is derived from the spectral position of the bare DSPP mode, as
shown in Fig.\,\ref{fig:W_Near}(b). The eigenfrequency of the bare
LSPP mode, $\omega_{\mathrm{L}}\left(E_{\mathrm{F}}\right)$, can be derived
from fitting the simulation result to the relation $\omega_{\mathrm{L}}\propto\sqrt{E_{\mathrm{F}}}$
\cite{nikitin2012surface,ju2011graphene}. Eventually, the eigenfrequency
of the induced polariton modes is found as \cite{salomon2012strong,agranovich2003cavity}

\begin{equation}
\omega_{\mathrm{u,l}}\left(E_{\mathrm{F}}\right)=\frac{\omega_{\mathrm{L}}\left(E_{\mathrm{F}}\right)+\omega_{\mathrm{D}}}{2}\pm\sqrt{\omega_{\delta}^{2}+\frac{\left[\omega_{\mathrm{L}}\left(E_{\mathrm{F}}\right)-\omega_{\mathrm{D}}\right]^{2}}{4}}.\label{eq:polariton_branch}
\end{equation}

The analytical results,depicted in Figs.\,\ref{fig:Fermi_Near}(c)
and (d) as blue solid curves,  are in excellent agreement with the simulated ones.

\section{Far-field interference scheme (FFIS)}

\begin{figure}
\begin{centering}
\includegraphics[width=84mm]{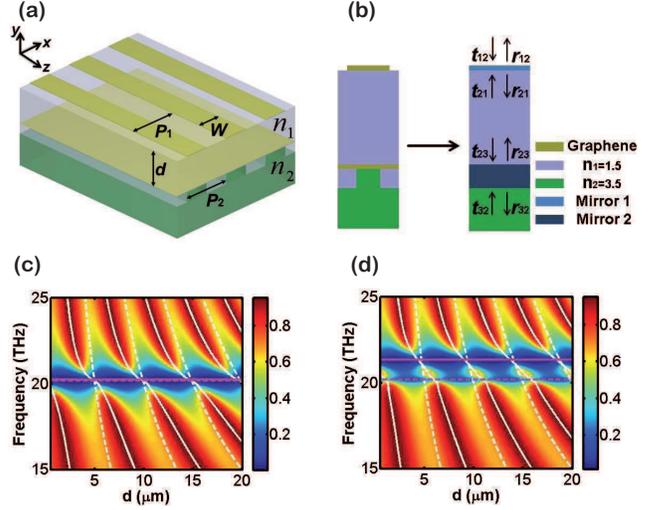}
\par\end{centering}

\caption{(a) Schematic view of the far-field coupling system. The duty cycle and the depth of the grating is fixed
to $0.5$ and $300\,\mathrm{nm}$, respectively. Note that the
individual graphene ribbon in the upper periodic array is exactly
on top of the corrugated part of the diffractive grating. (b) The
entire system shown in panel (a) can be effectively described as a
FP cavity with strongly dispersive resonant mirrors as long as the
near-field coupling effect is negligible. (c) and (d): Simulated transmission
contour plots at two different ribbon widths: (c) $W=240\,\mathrm{nm}$;
(d) $W=228\,\mathrm{nm}$. Bare FP modes are presented as the white
dashed lines, and the coupled modes computed from the semi-analytical
model (Eq.\,\ref{eq:res_con}) are depicted as white solid lines.
Magenta solid and dashed horizontal lines represent the spectral positions
of the bare LSPP and DSPP modes, respectively.\label{fig:Schematic_Far}}
\end{figure}

Using the NFCS as shown in Fig.\,\ref{fig:Schematic_Near}, it is
impossible to explore the far-field interaction between DSPPs
and LSPPs. When the separation distance between the upper PAGR
and the bottom graphene layer is well beyond the decay length of the evanescent diffraction orders,
the DSPP mode, supported by the bottom graphene layer, cannot be excited
at all. Thus, a newly proposed
structure, that we suggest to term far-field interference scheme (FFIS)  displayed
in Fig.\,\ref{fig:Schematic_Far}(a), is exploited to investigate
the far-field interaction between DSPPs and LSPPs in graphene.
It distinguishes from the NFCS in that a dielectric diffractive grating beneath the graphene
layer causes the excitation of the DSPP mode as suggested in Ref.\,\onlinecite{gao2012excitation},
even if the upper PAGR is far apart.

As displayed in Fig.\,\ref{fig:Schematic_Far}(a), the upper PAGR,
with array period $P_{1}$ and ribbon width $W$, and the lower graphene
layer are separated by a dielectric spacer ($\varepsilon_{1}=2.25$) of thickness
$d$. Here we use a silicon ($\varepsilon_{2}=12.25$) diffractive grating with
grating period $P_{2}$ underneath the graphene layer to facilitate
the excitation. For simplicity, we set $P_{1}=P_{2}=P$. It is evident that in the FFIS
near-field coupling between
DSPP and LSPP modes is still present provided that the separation distance $d$
is small enough. However, here a sufficiently large separation distance will be assumed in order to exclusively focus
on the far-field interaction phenomenon between DSPPs and
LSPPs. In passing we mention that the near-field effect in the FFIS resembles that in the NFCS.

Considering the transition to the far-field interference, as the structure
dimension along the propagation direction of the incident
wave is in the order of the wavelength, retardation effects have to
be taken into account. In addition to that, pronounced Fabry-Perot
(FP) modes can be supported in the system for large separation distances
$d$. Thus, the FFIS can be effectively described as a FP cavity bounded
with two strongly dispersive resonant mirrors as shown in Fig.\,\ref{fig:Schematic_Far}(b).
Since the extension of the two effective resonant mirrors along the
wave propagation direction is extremely subwavelength, they can be
regarded as metasurfaces. Therefore, the resonance condition characterizing
the transmission maxima for the entire system reads

\begin{equation}
2n_{1}d\frac{\omega}{c}+\varphi_{1}\left(\omega\right)+\varphi_{2}\left(\omega\right)=2\pi M,\label{eq:res_con}
\end{equation}
where $c$ is the speed of light in vacuum, $n_1=\sqrt{\varepsilon_1}$ is the refractive index of the spacer, $M$ is a positive integer, $\varphi_{1}\left(\omega\right)=\arctan\left[\Im\left(r_{21}\right)/\Re\left(r_{21}\right)\right]$
and $\varphi_{2}\left(\omega\right)=\arctan\left[\Im\left(r_{23}\right)/\Re\left(r_{23}\right)\right]$
are the phase shifts upon reflection at the two effective resonant mirrors, respectively.
The reflection/transmission coefficients indicated in
Fig.\,\ref{fig:Schematic_Far}(b), such as $r_{21},t_{21}$ and $r_{23},t_{23}$,
can be straightforwardly numerically calculated for either isolated mirror. Moreover, the transmission
coefficient of the entire system for any specific larger separation
distance $d$ can be semi-analytically calculated using Airy's formula \cite{teich1991fundamentals}

\begin{equation}
t=\frac{t_{12}t_{23}\exp\left(i\varphi\right)}{1-r_{21}r_{23}\exp\left(i2\varphi\right)},\label{eq:t_ana}
\end{equation}
where $\varphi=n_{1}d\left(\omega/c\right)$.

Let us start
the analysis with $P=300\,\mathrm{nm}$, and the identical Fermi energy
$E_{\mathrm{F}}=600\,\mathrm{meV}$ in upper PAGR and
lower graphene layer. The simulated transmission spectra as a function of the spacer distance $d$
for two different ribbon widths $W$ are shown in Figs.\,\ref{fig:Schematic_Far}(c)
and (d). The spectral positions of the bare LSPP ($\omega_{\mathrm{L}}$)
and DSPP ($\omega_{\mathrm{D}}$) modes are displayed by magenta solid
and dashed lines, respectively, and the bare FP modes are shown
as white dashed lines. It is evident that the semi-analytical formula
(Eq.\,\ref{eq:res_con}) accurately reproduces the resonance positions
for larger separation distance $d$ as shown by the white solid lines.

In Fig.\,\ref{fig:Schematic_Far}(c)  the width $W=240\,\mathrm{nm}$ is chosen such that
 $\omega_{\mathrm{L}}\approx\omega_{\mathrm{D}}$.
In this scenario, the periodic transmittance pattern with transmission
dips around the magenta solid/dashed line indicates the formation of
hybrid  LSPP-DSPP-FP modes. The spectral line width of the hybrid
mode varies periodically, which can be attributed to the tailored
spatial arrangement of the plasmonic modes (LSPP and DSPP modes).
The spectral line width of the hybrid mode decreases
drastically at definite spacer thicknesses where the Bragg criterion
$d=N\pi n_{1}\omega_{\mathrm{L}}/c$ is fulfilled
\cite{taubert2011near}.

A more interesting scenario occurs for $W=228\,\mathrm{nm}$ where $\omega_{\mathrm{L}}\neq\omega_{\mathrm{D}}$,
as shown in Fig.\,\ref{fig:Schematic_Far}(d).
As a consequence of the spectrally detuned interactions between either
the LSPP or the DSPP mode with the FP modes, a sequence of high transmittance
spots arise between
the magenta solid and dashed lines. They can be attributed to the fact that the FP modes are modulated
by spectrally detuned LSPP and DSPP modes together. The transmission
dips around the black solid (dashed) line are mainly induced by the
formation of the hybrid LSPP-FP (DSPP-FP) modes. There
is an obvious spectral distinction between these two kinds of transmission
dips, as shown in Fig.\,\ref{fig:Schematic_Far}(d), which is mainly
due to the fact that the phase shift at the two effective mirrors is different.

\section{Conclusion}

In this work we proposed two graphene based systems to facilitate
the manipulation of the interaction between a localized surface
plasmon polariton (LSPP) and a delocalized surface plasmon polariton
(DSPP) in the near-field as well as in the far-field regime for the
first time. In the near-field regime, both LSPP and DSPP modes can
be regarded as particle-like plasmonic oscillators in the framework
of plasmon hybridization theory. Therefore,
by changing the geometrical parameters in the near-field coupling
scheme (NFCS) and the Fermi energy of graphene, one can actually manipulate
the physical properties of the individual oscillators. This allows to control the
interaction between them as an entity.  However,
in the far-field regime, the radiative interaction mediated by the Fabry-Perot
(FP) modes supported by the far-field interference scheme (FFIS) will
be more pronounced. Thus, a picture of an effective FP cavity is proposed
to provide an intuitive understanding for this interaction regime. Our
findings may open up an avenue for the development of compact elements
such as tunable sensors, switchers, and filters at IR and THz frequencies.

\section*{Acknowledgements}
This work was supported by the German Federal Ministry of Education and Research (PhoNa) and by the Thuringian State Government (MeMa).


\end{document}